\documentclass{elsart}
\journal{Phys. Lett. B}
\usepackage{amsmath,graphicx,amssymb}
\newcommand{\be}{\nuc{8}{B}}
\newcommand{\uC}{\ensuremath{\mathrm{C}}}  
\newcommand{\uE}{\ensuremath{\mathrm{E}}}  
\newcommand{\ud}{\mathrm{d}}
\newcommand{\urms}{\ensuremath{\mathrm{rms}}}
\newcommand{\dbde}{\ensuremath{\frac{\mathrm{d} B (\mathrm{E}\lambda)}
{\mathrm{d}E}}}                            
\begin{document}
\begin{frontmatter}
\title{Analytical approach to electromagnetic processes in
loosely bound nuclei: application to \nuc{8}{B}}
\author[chalmers]{C.~Forss\'en\corauthref{cor}},
\corauth[cor]{Corresponding author.}
\ead{c.forssen@fy.chalmers.se}
\author[chalmers,kurchatov]{N.~B.~Shul'gina} and
\author[chalmers]{M.~V.~Zhukov}
\address[chalmers]{Department of Physics, Chalmers University of
Technology and G{\"o}teborg University, S--412~96 G{\"o}teborg, Sweden}
\address[kurchatov]{The Kurchatov Institute, 123182 Moscow, Russia}
\begin{abstract}
In this paper we develop an analytical model in order to study
electromagnetic processes involving loosely bound neutron--rich and
proton--rich nuclei. We construct a model wave function, to describe
loosely bound few--body systems, having the correct behaviour both at
large and small distances. The continuum states are approximated by
regular Coulomb functions. As a test case we consider the two--body
Coulomb dissociation of \nuc{8}{B} and, the inverse, radiative capture
reaction. The difference between using a pure two--body model and the
results obtained when incorporating many--body effects, is
investigated. We conclude that the interpretation of experimental data
is highly model dependent and stress the importance of measuring
few--body channels.
\end{abstract}
\end{frontmatter}
Electromagnetic processes such as radiative capture, photo--dissociation
and Coulomb dissociation have always been, and still remain, an
excellent tool to investigate nuclear structure. Unfortunately, for
radioactive nuclei the radiative capture experiments are very difficult
and studies of photo--dissociation are virtually impossible. However,
with the advent of radioactive beam facilities, nuclear structure of
dripline nuclei can be studied using Coulomb dissociation on heavy
targets. The Coulomb breakup of these nuclei is of interest also in
nuclear astrophysics, since it can be related to the corresponding
radiative capture process at astrophysical energies~\cite{smi01:51}. In
this paper, we will present an analytical model for Coulomb dissociation
(and consequently for radiative capture) of loosely bound dripline
nuclei (see also our recent conference
contribution~\cite{for02:unp}). We want to stress that our approach will
be general in the sense that both neutron--rich, and proton--rich
systems can be studied. The possibility to study reactions analytically
is very appealing since it often allows a deeper physical understanding
of the process. For some cases such studies are not only possible but
might even give results that are directly comparable to experimental
data and more advanced, numerical calculations. In particular, this is
the case for electromagnetic transitions between a loosely bound nuclear
state and a pure Coulomb continuum. ``Loosely bound'' implies that the
nucleus will exhibit a large degree of clusterization. We will consider
the case of two clusters, but will also discuss the three--cluster
channel (see also similar approaches for
one--neutron~\cite{ber88:480,ots94:49,kal96:22} and
two--neutron~\cite{pus96:22,for02:697,for02:706} halo nuclei). After a
general description of our model we will exemplify it with an
application to \be. Our main focus will be the nuclear structure
effects, and we will utilize the advanced three--body model of
Grigorenko~et~al.~\cite{gri99:60}, while the Coulomb dissociation is
considered only to first order. As it was shown
recently~\cite{typ97:613,ban02:65}, higher order effects should not be
significant at beam energies higher than around 70 MeV/$A$.

Our starting point for calculating electromagnetic cross sections will
be the $\uE\lambda$ strength function for a transition from an initial
state to a final, continuum state with energy $E$
\begin{equation}
  \dbde=\frac{1}{2J_i + 1}
  \sum_{f} \int \d \tau_f\left| \langle f||
 \mathcal{M}(\uE \lambda) || i \rangle \right|^2
  \delta\left(E_f-E\right),
\label{eq:strength}
\end{equation}
where $\ud \tau_f$ is the phase space element for final states,
$\mathcal{M} (\uE \lambda)$ is the electric multipole operator and
$| i \rangle$, $| f \rangle$ are the initial and final states in the
center of mass subsystem.

We will consider loosely bound systems of two clusters $(c + x)$ and, in
particular, we will study transitions to the low--energy continuum in
which excitations are manifested as relative motion between the clusters
$E = \hbar^2 k^2 / 2 \mu_{cx}$, where $\mu_{cx}$ is the reduced mass of
the system. Introducing the intercluster distance, $r$, the
corresponding cluster $\uE \lambda$ operator (operating only on the
relative motion of clusters) is
\begin{equation}
  \mathcal{M}(\uE \lambda,\mu) = e Z(\lambda) r^\lambda Y_{\lambda \mu}
  (\hat{r}),
\label{eq:clusterel}
\end{equation}
with the effective multipole charge $Z(\lambda) = \mu_{cx}^\lambda ( Z_x /
m_x^\lambda + (-1)^\lambda Z_c / m_c^\lambda)$.

The strength function is the key to study several reactions. The
photo--dissociation $(A + \gamma \rightarrow c + x)$ cross section is
given by
\begin{equation}
  \sigma_\gamma^{\uE\lambda} (E_\gamma) = \frac{(2\pi)^3 (\lambda +
  1)}{\lambda [ (2\lambda + 1)!!]^2} \left( \frac{E_\gamma}{\hbar c}
  \right)^{2\lambda -1} \dbde,
\label{eq:xsecgamma}
\end{equation}
where the photon energy, $E_\gamma = E + E_0$, is larger than the
binding energy, $E_0$. The inverse radiative capture reaction can be
studied using detailed balance
\begin{equation}
  \sigma_\mathrm{rc}^{\uE\lambda} (E) = \left(
  \frac{E_\gamma}{\hbar c k} \right)^2 \frac{2 (2J_A +1)}{(2J_c + 1)
  (2J_x + 1)} \sigma_\gamma^{\uE\lambda} (E_\gamma),
\label{eq:xsecrc}
\end{equation}
where $J_i$ is the spin of particle $i$. Using first order perturbation
theory, and the method of virtual quanta~\cite{win79:319,ber85:442}, the
energy spectrum for Coulomb dissociation on a high--$Z$ target can be
written as a sum over multipole, $\pi \lambda$, photo--dissociation
cross sections multiplied by the corresponding spectra of virtual
photons, $n_{\pi \lambda} (E_\gamma)$,
\begin{equation}
  \frac{\ud \sigma_\uC}{\ud E} = \sum_{\uE\lambda} \frac{n_{\uE
  \lambda} (E_\gamma)}{E_\gamma} \sigma_\gamma^{\uE\lambda} (E_\gamma) +
  \sum_{\mathrm{M}\lambda} \frac{n_{\mathrm{M} \lambda}
  (E_\gamma)}{E_\gamma} \sigma_\gamma^{\mathrm{M}\lambda} (E_\gamma).
\label{eq:xsecemd}
\end{equation}
Note that since M$\lambda$ transitions are usually~\cite{ber85:442}
strongly suppressed we will not consider them in this work.
In order to study all these reactions analytically we will propose a
model function to describe the (loosely) bound state of a two--body
system. We will only be interested in direct transitions to a ``clean''
continuum, i.e., with all nuclear phase shifts equal to zero. Thus, the
final state, with angular momentum $l_f$ between the clusters, will be
described by a regular Coulomb function
\begin{align}
  \phi_{l_f} (k,r) &= \sqrt{\frac{2}{\pi}} \frac{1}{k} i^{l_f} e^{i
  \sigma_{l_f}} F_{l_f} (k,r), \qquad \mathrm{where} \\
  F_{l_f} (k,r) &= C_{l_f}(k) e^{ikr} (kr)^{l_f + 1}
  {}_1F_1(l_f+1+i\eta (k) ; 2l_f + 2 ; -2ikr),
\label{eq:coulombwave}
\end{align}
and $\sigma_{l_f}$ is the Coulomb phase, $\eta (k) = Z_c Z_x e^2
\mu_{cx} / \hbar^2 k$ is the Sommerfeld parameter, $C_{l_f} (k) =
2^{l_f} e^{-\pi \eta (k) / 2} |\Gamma(l_f + 1 + i\eta (k))| / (2l_f +
1)!$, and ${}_1F_1(a;b;z)$ is the confluent hypergeometric
function~\cite{abr72}.

The reduced matrix element introduced in the definition of the strength
function, Eq.~\eqref{eq:strength}, contains a radial integral. With our
approximation for the continuum state this integral takes the form
\begin{equation}
  I_{l_f} (k) = \int_0^\infty \ud r e^{-i k r} r^{l_f + 1} {}_1F_1 (l_f
  + 1 - i \eta (k) ; 2l_f + 2 ; 2 i k r) r^\lambda \phi_b (r).
\label{eq:radint}
\end{equation}
Here, $\phi_b (r)$ is the two--body, relative motion wave function (WF)
describing the initial state.  For large $r$, and with angular momentum
$l_i$ between the clusters, this radial function should be proportional
to the Whittaker function $W_{-\eta_0,l_i+1/2} (2 \kappa_0 r)$ (see
Ref.~\cite{abr72}), where $\eta_0 = Z_c Z_x e^2 \mu_{cx} / \hbar^2
\kappa_0$ and $E_0 = \hbar^2 \kappa_0^2 / 2 \mu_{cx}$ is the binding
energy.

In most studies on loosely bound systems, the Whittaker function has
been used to describe the bound state for all $r$. However, this
approximation is only motivated if the transition matrix element is
dominated by contributions from very large $r$. This should be the case
for reactions at very small energies. For real experimental energies ($E
\gtrsim 100$~keV), the WF of the bound state should be constructed in a
more realistic way. Therefore we will introduce a model function that
describes the bound state $c + x$ WF accurately for all distances. This
is done by considering the behavior at small and large $r$. We have
already pointed out the the WF should be described by a Whittaker
function at large $r$. Furthermore, for a two--body system, consisting
of point--like particles, it should behave as $r^{l_i + 1}$ as $r
\rightarrow 0$. Both asymptotics are fulfilled using the following model
function
\begin{equation}
  \phi_{b,\tilde{\gamma}}^\mathrm{``exact''} (r) =
  \frac{1}{\sqrt{N_{\tilde{\gamma}}}}
  W_{-\eta_0,l_i+1/2} (2 \kappa_0 r) \left(1 - e^{-\kappa_1 r}
  \right)^{2l_i + 1},
\label{eq:exactmodelwf}
\end{equation}
where $N_{\tilde{\gamma}}$ is the normalization constant and
$\tilde{\gamma}$ denotes the parameters $\left\{ \kappa_0, \kappa_1,
\eta_0 \right\}$. The parameters $\kappa_0$ and $\eta_0$ are defined by
the binding energy, charges and masses, while $\kappa_1$ can be fitted
to give the correct size of the system. Using this WF, and solving the
integral~\eqref{eq:radint} numerically, it is possible to get very good
estimates for the electromagnetic reaction cross sections. As a remark
we want to point out that, for a one--neutron halo nucleus ($\eta_0 =
0$), the Whittaker function will transform into a modified, spherical
Bessel function: $W_{0,l_i+1/2} (2 \kappa_0 r) = \sqrt{2 \kappa_0 r /
\pi} K_{l_i + 1/2} (\kappa_0 r)$. In this case, the
integral~\eqref{eq:radint} can be solved exactly.

However, we are searching for a completely analytical model which will
also enable us to incorporate many--body effects. Our model
function~\eqref{eq:exactmodelwf} has to be modified accordingly.  First,
we note that the asymptotic form of the Whittaker function as $r
\rightarrow \infty$ is
\begin{equation}
  W_{-\eta_0,l_i+1/2} (2 \kappa_0 r) \sim e^{-\kappa_0 r} r^{-\eta_0}
  \left[ 1 + \mathcal{O} (1/r) \right].
\label{eq:asympwhit}
\end{equation}
Secondly, for two--body systems in which the clusters have an internal
structure, the centrifugal barrier is effectively larger and the WF
should behave as $r^n$, where $n > l_i + 1$, as $r \rightarrow 0$.

Motivated by this, we put forward the following model function
\begin{equation}
\begin{aligned}
  \phi_{b,\gamma} (r) &= \frac{1}{\sqrt{N_\gamma}}
  \frac{e^{-\kappa_0 r}}{r^{\eta_0'}} \left(1 - e^{-\kappa_1 r}
  \right)^p, \qquad \mathrm{with~norm}\\
  N_\gamma &= \sum_{m=0}^{2p} \binom{2p}{m} (-1)^m
  (2\kappa_0 + m\kappa_1)^{2\eta_0' - 1} \Gamma(1 - 2\eta_0'),
\end{aligned}
\label{eq:modelwf}
\end{equation}
where $\gamma$ denotes the parameters $\left\{ \kappa_0, \kappa_1,
\eta_0', p \right\}$. The parameter $\kappa_0$ is defined by the binding
energy and effective mass. By putting $\eta_0' = \eta_0$ we would ensure
to reproduce the tail of the WF at very large $r$. However, higher order
terms in Eq.~\eqref{eq:asympwhit} remain important for $r \lesssim
100$~fm. Therefore, $\eta_0'$ and $\kappa_1$ are used as free parameters
in a fit to the ``exact'' WF~\eqref{eq:exactmodelwf} in the interval of
interest. In this way $\eta_0'$ will be an ``effective'' Sommerfeld
parameter while $\kappa_1$ will still mainly be connected with the
size. Finally, the integer $p$ is fixed by the small $r$ behaviour,
$r^n$, and will thus allow us to take many--body effects into account.

It is now possible to make the integral~\eqref{eq:radint} analytically
(see Ref.~\cite{gra80})
\begin{equation}
\begin{split}
  I_{l_f,\gamma} (k) = \frac{1}{\sqrt{N_\gamma}} \sum_{m=0}^{p}
  \binom{p}{m} (-1)^m (m \kappa_1 + \kappa_0 + i k)^{-(l_f + 2 + \lambda
  - \eta_0')} \Gamma (l_f + 2 + \lambda - \eta_0')\\ \times {}_2F_1
  \left( l_f + 2 + \lambda - \eta_0' ; l_f + 1 - i \eta (k) ; 2 l_f + 2
  ; \frac{2 i k}{m\kappa_1 + \kappa_0 + i k} \right).
\label{eq;anaint}
\end{split}
\end{equation}
Many-body nuclear structure can further be taken into account by
considering the possibility that the bound state WF contains several
different two--body components
\begin{equation}
  \phi_b (r) = \sum_i a_i \phi_{b,\gamma_i} (r).
\label{eq:comp}
\end{equation}
Note that pure many--body components will not contribute to two--body
break\-up, but will instead lead to $\sum_i a_i^2 < 1$. Note also that the
threshold for two--body breakup will be higher for components where one
(or both) of the clusters is excited. Therefore, we define the continuum
strength function separately for each component. Finally, we arrive at
an analytical formula for the strength function
\begin{equation}
  \left. \dbde \right|_i = \frac{e^2 Z^2(\lambda) \mu_{cx}}{\hbar^2}
  \frac{2\lambda + 1}{2\pi^2} \sum_{l_f} a_i^2 k^{2 l_f + 1} C_{l_f}^2
  (k) \langle l_i 0 \lambda 0 | l_f 0 \rangle^2 | I_{l_f,\gamma_i} (k)
  |^2.
\label{eq:anastrength}
\end{equation}
%

As an example we will apply our analytical approach to the $^{8}$B
nucleus. Our key point here will be to show the applicability of our
analytical model and to demonstrate the importance of many--body nuclear
structure. The interest in $^{8}$B stems from its key role in the
production of high-energy solar neutrinos. The probability of the
reaction \nuc{7}{Be}$ + p \rightarrow$\be $+ \gamma$ at solar energies
strongly depends on the structure of \be\ and, in particular, on the
asymptotics of the valence proton WF. The reaction can be studied
indirectly through Coulomb dissociation, using a radioactive $^{8}$B
beam on a heavy target~\cite{cor02:529,dav01:63,iwa99:83,kik97:391}. We
should also mention the recent progress in radiative capture
measurements~\cite{ham01:86}, where the cross section has been measured
at energies around 200 keV with an accuracy of $\approx 15$
percent. However, in all cases theoretical models are needed to
extrapolate the measured cross sections down to solar energies.

The low--lying \be\ continuum can, with relatively good precision, be
approximated as a pure Coulomb one. There are no negative parity states
at low excitation energies~\cite{gol98:67} and the electromagnetic
processes are, in all cases we are considering, dominated by $\uE 1$
transitions.

\begin{figure}[hbt]
  \includegraphics[width=0.75\textwidth]{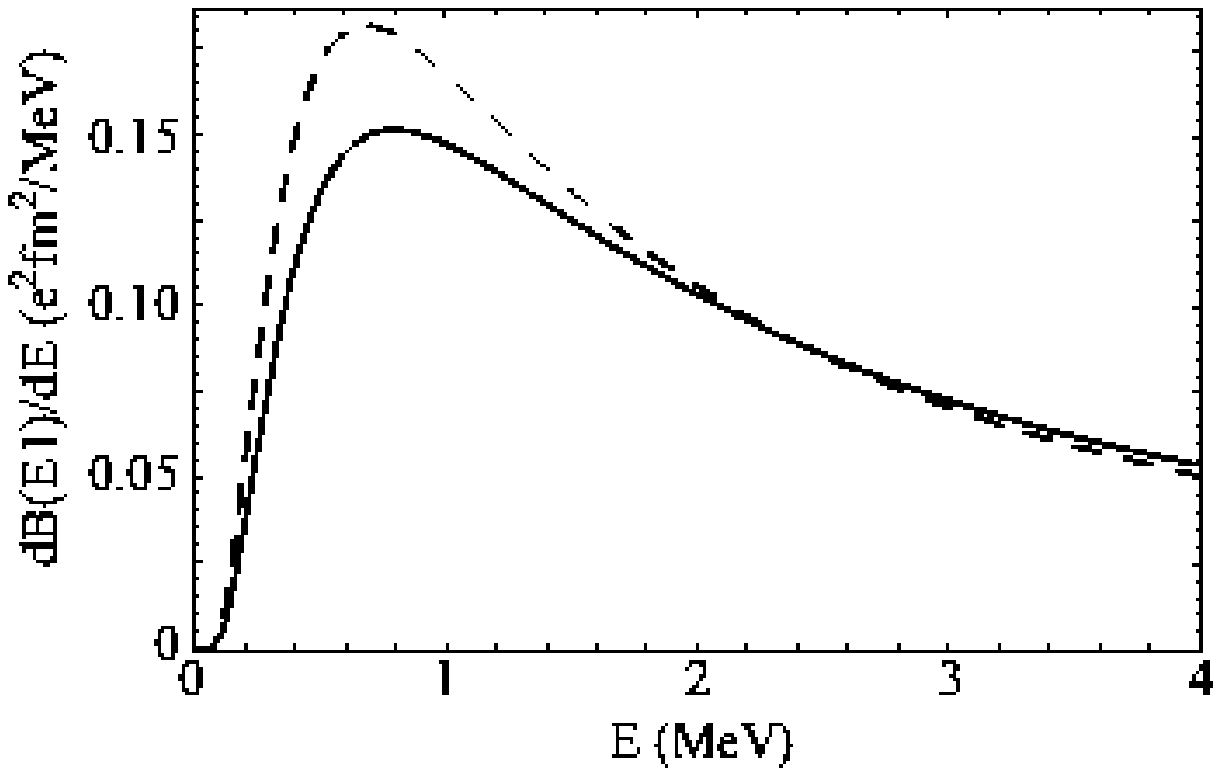}
  \caption{The E1 strength function of \be. Although the total strength
  is the same, the shapes are very different when treating the \be\
  nucleus as a two--body (dashed line) or a three--body (solid line)
  system. The difference is mainly due to the stronger centrifugal
  barrier in the three--body case forcing the WF to be narrower and thus
  wider in momentum/energy space. The parameters of the model WFs can be
  found in Table~\ref{tab:parameters}.  %
  \label{fig:e1strength}}
\end{figure}

We start by treating \be\ as a pure two--body ($\nuc{7}{Be} + p$) system
with binding energy $E_0 = 137$~keV and relative orbital momentum $l_i =
1$. The single free parameter, $\kappa_1$, in our ``exact'' model
function~\eqref{eq:exactmodelwf} is then fitted to an rms intercluster
distance of $r_\urms = 4.57$~fm (extracted from
Ref.~\cite{gri99:60}). In order to get analytical results, we then
introduce the model function~\eqref{eq:modelwf}. With $p = 2l_i + 1 = 3$
and $\kappa_0$ fixed from the binding energy, the remaining parameters
$\kappa_1$ and $\eta_0'$ are fitted to the behavior of the ``exact''
model function~\eqref{eq:exactmodelwf}, see
Table~\ref{tab:parameters}. The resulting E1 strength function is showed
as a dashed line in Fig.~\ref{fig:e1strength}. This analytical
approximation agrees, to a very high precision, with the numerical
results obtained keeping the ``exact'' model function. The error is less
than 2~\% in the region of interest.

However, concerning the structure of the \be\ ground state, one should
keep in mind that the \nuc{7}{Be} core is in itself a weakly bound
system with an excited $1/2^-$ state at 429~keV. The common treatment of
\be\ as a two--body system is therefore highly questionable. We now want
to investigate what effect the many--body structure of \be\ might have
on the strength function. We utilize a recent three--body
calculation~\cite{gri99:60}, where it was shown that, after projection
on the two--body channel, there are three main components (adding up to
94~\% of the total WF), and that the rest are pure three--body channels,
see Table~\ref{tab:parameters}. For each of these two--body components
we fit our parameters $\kappa_1$ and $\eta_0'$. The binding energy, $E_0
= 137$~keV, determines $\kappa_0$ for the two first components and $E_0
= 566$~keV for the third, \nuc{7}{Be} excited state, component. The best
fit of the small $r$ behavior is obtained with $p = 5$ which reflects
the effectively larger centrifugal barrier in the three--body case. This
centrifugal barrier will push the WF away from $r=0$ and will, thus,
force it to become more narrow than the two--body WF. We therefore
expect the distribution in momentum/energy space to be broader. This
effect is clearly seen in Fig.~\ref{fig:e1strength} where the E1
strength function obtained using this three--body model\footnote{Note
that this is not strictly a three--body model, but rather the two--body
projection of a three--body WF. However, in the following we will
consistently refer to it as three--body results.} is shown as a solid
line. This difference, seen in the strength function, should be even
more pronounced in the energy spectrum. From these results one can
conclude that the interpretation of energy spectra is highly model
dependent.

\begin{figure}[hbt]
  \begin{minipage}{0.95\textwidth}
    \begin{minipage}[t]{0.47\textwidth}
      \centering
      \includegraphics[width=\textwidth]{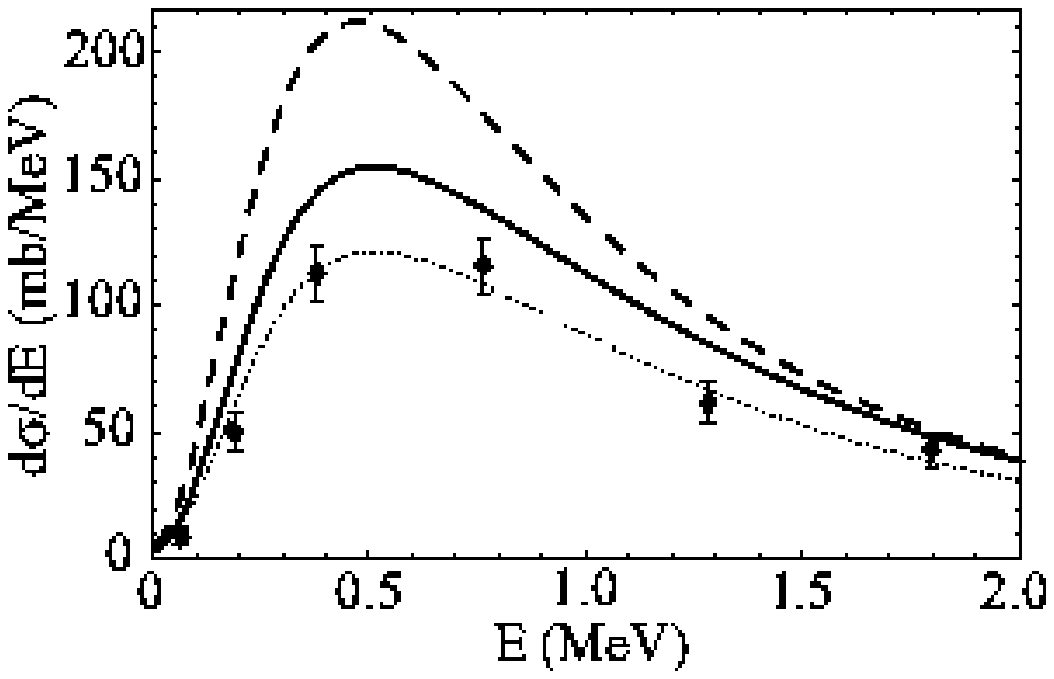}
    \end{minipage}
  \hfill
    \begin{minipage}[t]{0.47\textwidth}
      \centering
      \includegraphics[width=\textwidth]{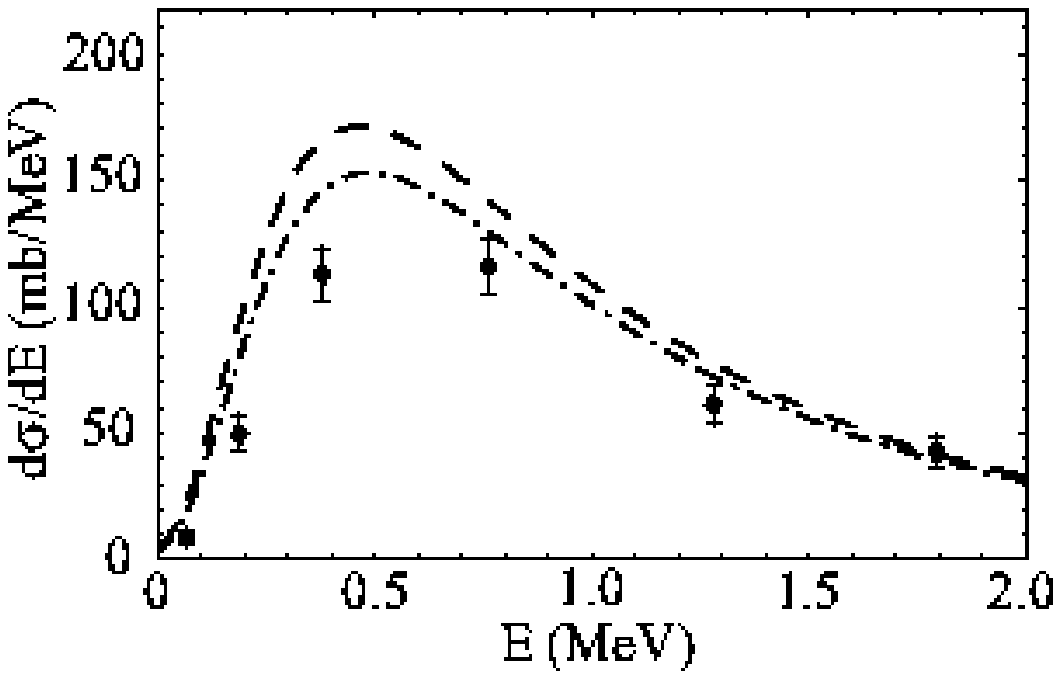}
    \end{minipage}
    \begin{minipage}[b]{0.47\textwidth}
      \centering
      (a)
    \end{minipage}
  \hfill
    \begin{minipage}[b]{0.47\textwidth}
      \centering
      (b)
    \end{minipage}
  \end{minipage}
  \caption{The \be\ Coulomb dissociation energy spectrum obtained at
  82.7~MeV/$A$ on Pb with \be\ scattering angles $\leq 1.77^\circ$. The
  data points are from~\cite{dav01:63}. In (a) the curves show our
  analytical two--body (dashed line) and three--body (solid line)
  calculations, while the thin, dotted line is just a scaled version of the
  three--body result. In (b) the dash--dotted line is the cross section
  calculated numerically within a two--body potential
  model, from~\cite{ber96:365}. A comparison is made with our analytical
  two--body model (dashed line) having the same intercluster
  distance. All theoretical curves have been corrected for
  experimental resolution and acceptance.%
  \label{fig:dsde}}
\end{figure}

In Fig.~\ref{fig:dsde}(a) we compare our analytical results for Coulomb
dissociation, including both E1 and E2 transitions, to the experimental
data from Davids~{et~al.}~\cite{dav01:63}. This experiment is very
appealing since the selection of scattering angles ($\theta_{\be} \leq
1.77^\circ$) minimizes the contribution from nuclear scattering and the
relatively high beam energy (82.7~MeV/$A$) justifies the use of first
order perturbation theory. Concerning the shape of the energy spectrum
we have an excellent agreement between the experimental data and our
results obtained using the three--body model (see thin, dotted line),
while the pure two--body calculation gives a too narrow peak. As to the
absolute values, the three--body model gives about 20~\% larger cross
section than the experimental data. In Fig.~\ref{fig:dsde}(b) we compare
the results from our two--body model (dashed line) to the results from a
potential model calculation without final state interaction,
from~\cite{ber96:365} (dash--dotted). Here, the intercluster distance
for both models is $r_\urms = 4.23$~fm which is lower than before
($r_\urms = 4.57$~fm). The lesson from this figure is twofold. Firstly,
it demonstrates the sensitivity to the intercluster distance and the
importance of knowing this parameter. Secondly, we note that the
difference between the two--body potential model calculation and our
analytical results is not so large. The results deviate mainly in the
peak amplitude.

\begin{figure}[hbt]
  \includegraphics[width=0.75\textwidth]{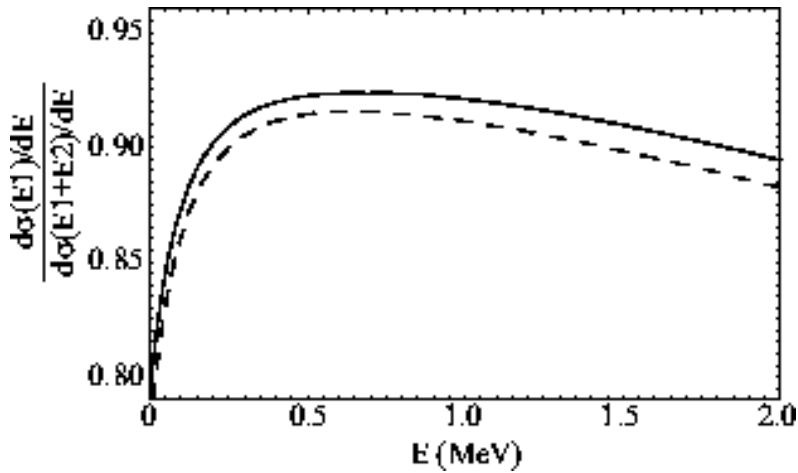}
  \caption{Fraction of the calculated cross section for the Coulomb
  dissociation of 82.7~MeV/$A$ \be\ on Pb with scattering angles $\leq
  1.77^\circ$ attributed to E1 transitions. The curves are two--body
  (dashed line) and three--body (solid line) calculations.%
    \label{fig:dsderatio}}
\end{figure}

In connection to the same experimental conditions we show in
Fig.~\ref{fig:dsderatio} the fraction of the cross section that can be
attributed to E1 transitions. This information is of importance when
studying the inverse radiative capture reaction and, in contrast to the
conclusion drawn by Davids~{et~al.}~\cite{dav01:63}, we claim that
although the E2 contribution increases for low relative energies, it
still does not dominate. Note that including M1 transitions will give
raise to a sharp dip at 640~keV.

We have also compared our results for the radiative capture reaction to
the experimental data by Hammache~{et~al.}~\cite{ham01:86}. Our
three--body result of $\sigma_\mathrm{rc} (186\mathrm{~keV}) = 16.1$~nb
agrees very well with the experimental value $16.7 \pm 2.1$~nb. We also
find that the E2 contribution to this cross section is around 0.03~\%.

In conclusion, we have performed analytical studies of electromagnetic
pro\-cesses for loosely bound nuclei. This has been accomplished by
using model radial functions that describe two--body WFs or the
two--body projections of many--body WFs, accurately for all radii, and
by only studying direct transitions to/from a pure Coulomb continuum. We
have examined the difference between a pure two--body model and the
results obtained incorporating many--body effects. From this we
concluded that the interpretation of experimental data is highly model
dependent. Comparisons have also been made to experimental results on
\be\ Coulomb dissociation and, the inverse, radiative capture
reaction. We found that our three--body results coincide with the
radiative capture cross section measured by
Hammache~{et~al.}~\cite{ham01:86} while our Coulomb dissociation energy
spectrum is about 20~\% larger than the experimental data from
Davids~{et~al.}~\cite{dav01:63}. However, the shapes of the experimental
and theoretical energy spectra are in excellent agreement. In contrast,
our pure two--body model of \be, having the same intercluster distance
$r_\urms$, does not agree with the experimental data. A final word of
wisdom is that, in order to interpret these data correctly, it is very
important to fix the spectroscopic factors of different two--body and
many--body components. Therefore, we want to stress the usefulness of
experiments where Coulomb dissociation is studied in complete
kinematics. Examples of interesting channels in the \be\ case is
\be$\rightarrow$\nuc{7}{Be}($1/2^-$)$+ p + \gamma$ and
\be$\rightarrow$\nuc{3}{He}$+ \alpha + p$. Recently, \nuc{7}{Be}
fragments and $\gamma$--rays were measured in coincidence after breakup
on a light target by Cortina--Gil~{et~al.}~\cite{cor02:529} and the
excited core component of the WF was clearly observed.

N.~B.~S.~is grateful for support from the Royal Swedish Academy of
Science. The support from RFBR Grants No 00--15--96590, 02--02--16174
are also acknowledged.
%

%
\newpage
%
\begin{table}[hbt]
\begin{center}
  \caption{Parameters of our model WFs used to describe the \be\ ground
  state. Both models give the same intercluster distance, $r_\urms =
  4.57$~fm. With $E_0 = 137$~keV we get $\kappa_0 = 0.076$~fm$^{-1}$ and
  $\eta_0 = 1.595$. The excited core component (last row) has $E_0 =
  566$~keV giving $\kappa_0 = 0.154$~fm$^{-1}$ and $\eta_0 = 0.786$. The
  relative orbital momentum for all components are $l_i =1$ while $I$ is
  the channel spin and $a^2$ is the spectroscopic factor. Note that
  there is no dependence on channel spin in our strength
  function~\eqref{eq:anastrength}.}
  \vspace{1ex}
  \begin{minipage}{\textwidth}
\begin{tabular}{l | l c c c c c}
    \hline
    \hline
      Model WF & configuration & $I$ & $a^2$ & $p$ & $\kappa_1$~(fm$^{-1}$) &
      $\eta_0' / \eta_0$ \\
	\hline
      	two--body & $\left[ \nuc{7}{Be}(3/2^-) \otimes p \right]$ & 2 &
    	1.00 & 3 & 0.601 & 0.79 \\
	\hline
	& $\left[ \nuc{7}{Be}(3/2^-) \otimes p \right]$ & 2 & 0.65 & 5 &
    	0.702 & 0.87 \\
	& $\left[ \nuc{7}{Be}(3/2^-) \otimes p \right]$ & 1 & 0.13 & 5 &
    	0.765 & 0.86 \\
	\raisebox{6ex}[0pt]{three--body} & $\left[ \nuc{7}{Be}(1/2^-)
    	\otimes p \right]$ & 1 & 0.16 & 5 & 0.753 & 1.43 \\
    \hline
    \hline
  \end{tabular}
  \end{minipage}
  \label{tab:parameters}
\end{center}
\vspace{-2.5mm}
\end{table}
%
\end{document}